\documentclass[twocolumn,aps,prl,superscriptaddress,showpacs,secnumroman,showkeys]{revtex4}
\usepackage{amsmath,bm}%
\usepackage{graphicx}%
\begin{document}
\title{Critical behavior of the isotope yield distributions in the Multifragmentation Regime of Heavy Ion Reactions}  
\author{M. Huang}
\affiliation{Institute of Modern Physics, Chinese Academy of Sciences, Lanzhou, 730000,China.}
\affiliation{Graduate University of Chinese Academy of Sciences, Beijing, 100049, China.}
\affiliation{Cyclotron Institute, Texas A$\&$M University, College Station, Texas, 77843, USA}
\author{R. Wada}
\email[E-mail at:]{wada@comp.tamu.edu}
\affiliation{Cyclotron Institute, Texas A$\&$M University, College Station, Texas, 77843, USA}
\author{Z. Chen}
\affiliation{Cyclotron Institute, Texas A$\&$M University, College Station, Texas, 77843, USA}
\affiliation{Institute of Modern Physics, Chinese Academy of Sciences, Lanzhou, 730000,China.}
\author{T. Keutgen}
\affiliation{FNRS and IPN, Universit\'e Catholique de Louvain, B-1348 Louvain-Neuve, Belgium}
\author{S. Kowalski}
\affiliation{Institute of Physics, Silesia University, Katowice, Poland.}
\author{K. Hagel}
\affiliation{Cyclotron Institute, Texas A$\&$M University, College Station, Texas, 77843, USA}
\author{M.Barbui}
\affiliation{Cyclotron Institute, Texas A$\&$M University, College Station, Texas, 77843, USA}
\author{A. Bonasera}
\affiliation{Cyclotron Institute, Texas A$\&$M University, College Station, Texas, 77843, USA}
\affiliation{Laboratori Nazionali del Sud, INFN,via Santa Sofia, 62, 95123 Catania, Italy}
\author{C.Bottosso}
\affiliation{Cyclotron Institute, Texas A$\&$M University, College Station, Texas, 77843, USA}
\author{T. Materna}
\affiliation{Cyclotron Institute, Texas A$\&$M University, College Station, Texas, 77843, USA}
\author{J. B. Natowitz}
\affiliation{Cyclotron Institute, Texas A$\&$M University, College Station, Texas, 77843, USA}
\author{L. Qin}
\affiliation{Cyclotron Institute, Texas A$\&$M University, College Station, Texas, 77843, USA}
\author{M.R.D.Rodrigues}
\affiliation{Cyclotron Institute, Texas A$\&$M University, College Station, Texas, 77843, USA}
\author{P.K. Sahu}
\affiliation{Cyclotron Institute, Texas A$\&$M University, College Station, Texas, 77843, USA}
\author{K.J. Schmidt}
\affiliation{Cyclotron Institute, Texas A$\&$M University, College Station, Texas, 77843, USA}
\author{J. Wang}
\affiliation{Institute of Modern Physics, Chinese Academy of Sciences, Lanzhou, 730000,China.}

\date{\today}

\begin{abstract}
Isotope yields have been analyzed within the framework of a Modified Fisher Model to study the power law yield distribution of isotopes in the multifragmentation regime. Using the ratio of the mass dependent symmetry energy coefficient relative to the temperature, $a_{sym}/T$, extracted in previous work and that of the pairing term, $a_{p}/T$, extracted from this work, and assuming that both reflect secondary decay processes, the experimentally observed isotope yields have been corrected for these effects. 
For a given I = N - Z value, the corrected yields of isotopes relative to the yield of $^{12}C$ show a power law distribution, $Y(N,Z)/Y(^{12}C) \sim A^{-\tau}$, in the mass range of $1 \le A \le 30$ and the distributions are almost identical for the different reactions studied. The observed power law distributions change systematically when I of the isotopes changes and the extracted $\tau$ value decreases from 3.9 to 1.0 as I increases from -1 to 3. These observations are well reproduced by a simple de-excitation model, which the power law distribution of the primary isotopes is determined to $\tau^{prim} = 2.4 \pm 0.2$, suggesting that the disassembling system at the time of the fragment formation is indeed at or very near the critical point.  
\end{abstract}
 


\maketitle
 

In the early 80's, the Purdue Group demonstrated that the isotope yields of intermediate mass fragments (IMFs) produced in high energy proton-nucleus collisions at the Fermi Lab exhibit a power law distribution with a $\tau$ value of 2.64-2.65 ~\cite{Finn82,Minich82,Hirsch84}. This observation stimulated the studies of critical phenomena and phase transitions in nuclear matter. Work through the  90's to early 2000's is well summarized both from the experimental and theoretical side in Refs.~\cite{Bonasera00,Lopez06}. In the mid 90's to 2000's, the Berkeley Group, applying Fisher's droplet model concepts to the experiments performed by the EOS and ISIS collaborations, argued that the disassembling system does indeed show a critical behavior~\cite{Gilkes94,Elliott00,Elliott02,Elliott03}. They  extracted a $\tau$ value of 2.2$\pm$0.1 from  both experiments. 

In some recent papers we have revisited the question of critical behavior in multi-fragmentation reactions resulting from violent collisions of heavy nuclei in the Fermi energy domain\cite{Bonasera08,Huang10_3} and discussed experimental evidence for a nuclear phase transition driven by different concentrations of neutrons and protons. Different ratios of the neutron to proton concentrations lead to different critical points for the phase transition.

One of the complications in multifragmentation originates from secondary statistical decay process. When fragments are formed in a disassembling system, they are generally excited and most de-excite to the ground state by the time of detection~\cite{Marie98,Staszel01,Hudan03}. According to Ref.\cite{Hudan03}, the average parent of Z=10 fragments produced in the Xe+Sn reaction at 39 A MeV emits $\sim 5.5$ mass units as $\sim 1.75$ charged particles and an additional $\sim$ 4 mass units as neutrons. This secondary decay process significantly alters the fragment isotopic distribution. Studies using statistical decay codes also indicate that the primary fragment distributions are significantly modified during the secondary decay process~\cite{Tsang01,Botvina02}. Most multifragmentation models, statistical or dynamical, take this process
into account, but the magnitude of the change depends on the codes and results can vary significantly \cite{Tsang06}. In the analysis of the Purdue Group, the secondary decay process was not taken into account and data for 4 $\le A \le 12$ were excluded from the fit in determining the $\tau$ values. 
In the analysis of the Berkeley group, since no mass was identified in either of the experiments, the secondary effects were treated empirically. In their analysis, the mass of each isotope was calculated as $2Z(1+y(E^{*}/B_{f}))$, where E* and $B_{f}$ are the fragment excitation energy and ground state binding energy and y is a free parameter. The parameters were determined to establish the power law between the scaled cluster yield and the scaled temperature~\cite{Elliott02}.
   
In order to get direct insight into the nature of the disassembling system at the time of the fragment formation, it is preferable to determine the secondary effects experimentally in particle-fragment correlations and use that information to reconstruct the yields of primary isotopes. However this is not straight forward, since multiple fragments are generally produced in a reaction and light particles can be produced even before the formation of the fragments, and therefore the identification of the parent for detected light particles observed in coincidence with a fragment is not trivial~\cite{Marie98,Staszel01,Hudan03}. Further, neutrons are particularly difficult as the multiplicity of neutrons from secondary decay is typically a very small fraction of the total neutron multiplicity.

In this work we focus on an alternative method in which the observed isotope distributions are corrected for known secondary decay effects to extract information on the properties of the disassembling system at the time of fragment formation. The role of secondary decay effects in modifying the original fragment distribution, and in particular its effect on determination of the critical parameter  $\tau$ is elucidated. 


The experiment was performed at the K-500 superconducting cyclotron facility at Texas A$\&$M University. 40 A MeV of $^{64,70}$Zn and $^{64}$Ni beams were used to irradiate $^{58,64}$Ni, $^{112,124}$Sn, $^{197}$Au and $^{232}$Th targets. IMFs were measured at 20$^\circ$ and typically 6-8 isotopes for atomic numbers, Z, up to Z=18 were clearly identified. The yields of light charged particles (LCPs) in coincidence with IMFs were also measured using 16 single crystal CsI(Tl) detectors. The details of the data analysis and results can be found in refs.~\cite{Huang10_1,Chen10}  

\begin{figure}
\includegraphics[scale=0.4]{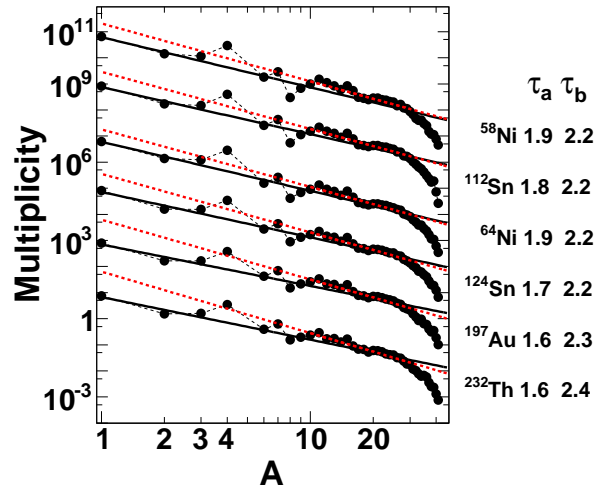}
\caption{\footnotesize 
(Color online) Experimental multiplicity distributions vs isotope mass, A, for the $^{64}$Ni projectile with different targets. Targets are indicated on the right for each distribution. Each data point represent summed multiplicity over Z for a give A. Solid lines are the resuts of power law fits for $1 \le A \le 30$ and dotted lines are for $10 < A \le 30$. The extracted $\tau$ values are also given as $\tau_{a}$ and $\tau_{b}$, respectively. 
}			
\label{fig:fig1_MvsA}
\end{figure}

In Fig.\ref{fig:fig1_MvsA}, the multiplicity distributions of the observed isotopes are plotted as a function of A for the case of the $^{64}$Ni projectile on different targets. The data are plotted from top to bottom as N/Z of the target increases. The distributions roughly show a power law distribution up to $A = 30$. Above $A = 30$, the multiplicity decreases sharply for all cases. In the figure the distributions are fit by a power law distribution, A$^{-\tau}$, in two different ranges of A, one (solid lines, $\tau_{a}$) is obtained with $1 \le A \le 30$ and the other (dotted lines, $\tau_{b}$) with $10 \le A \le 30$. In the latter cases the extracted values are in the range of 2.2 to 2.4, and slowly increase as N/Z of the target increases. The values are slightly smaller than those extracted by the Purdue group~\cite{Finn82,Minich82,Hirsch84} which were extracted from a similar range of A. On the other hand when the distributions are fit in a wider range extended to A=1 ,
  the extracted $\tau$ values becomes smaller ($1.6 \le \tau \le 1.9$), and decrease when the target N/Z increases. In both cases the data fluctuate along the fitted lines for smaller IMFs. $A = 4$ yields are always higher than the fit lines and $A = 8$ yields are significantly lower.
For other reaction systems, similar results are observed. These observations suggest that the secondary decay process plays a significant role in these distributions. To elucidate the role of the secondary decay process, the multiplicity deistributions are examined in detail, using information from the wide variety of isotopes identified in this experiment.

In the previous work of Ref.~\cite{Huang10_1}, we extracted the ratio between the symmetry energy coefficient and the temperature, $a_{sym}/T$, from the isobaric yield ratios of IMFs in a given reaction, based on the Modified Fisher Model~\cite{Minich82,Hirsch84}. 
In another work, Ref.~\cite{Chen10}, the $a_{sym}/T$ values are evaluated by two other independent methods. One uses isoscaling parameters determined from the ratio of the isotope yields between two different reactions. The other employs the variance of the isotope yield distribution in a single reaction.
All results from the three different methods are in reasonable agreement and indicate that the extracted values of $a_{sym}/T$ depend significantly on the mass number, A, of the fragment, i.e., $a_{sym}/T$ gradually increases from 4 $\sim$ 6 to 12 $\sim$ 16 as A increases from 9 to 37. These values depend slightly on the different methods, but the essential trends are quite similar. The extracted values can be empirically fit by 
\begin{eqnarray}
a_{sym}^{emp}/T &=& 5 + 1.4 (A - 9)^{\frac{2}{3}}\hspace{1.0cm} \textrm{for} A \ge 9 \nonumber\\
          &=& 5\hspace{3.2cm} \textrm{for} A < 9.
\label{eq:eq_asymEmp}
\end{eqnarray}

In those papers, detailed comparisons to AMD model simulations~\cite{Ono96,Ono99} incorporating a statistical decay code Gemini~\cite{Charity88} as an afterburner show that the experimentally observed A dependence is very well reproduced. In contrast the $a_{sym}/T$ values extracted from the primary isotope yield distributions of the AMD calculations, before cooling with the afterburner, are nearly constant with $a_{sym}/T \sim$ 4 to 6 (depending on extraction method) over the mass range of the observed isotopes, indicating that the experimentally observed A dependence of the symmetry energy term originates from the secondary statistical decay of the excited primary fragments.

 
The Modified Fisher Model of refs.~\cite{Minich82,Hirsch84} has been used to study the isotopic distributions of the fragments. In this model, the yield of A nucleons with I=N-Z, Y(A,I) is given by
\begin{eqnarray}
Y(A,I) &=& CA^{-\tau}exp\{[(F(A,I,T,\rho)+\mu_{n}N+ \mu_{p}Z)/T] \nonumber\\
&&+Nln(N/A)+Zln(Z/A)\},     
\label{eq:eq_yield}
\end{eqnarray}
where C is a constant. The A$^{-\tau}$ term originates from the entropy of the fragment. $\mu_{n}$ and $\mu_{p}$ are the neutron and proton chemical potentials, respectively. The last two terms are from the entropy of mixing of neutrons and protons \cite{Fisher67}. F(A,I,T,$\rho$) is the free energy of the cluster at temperature T and density $\rho$. 

\begin{figure}
\includegraphics[scale=0.4]{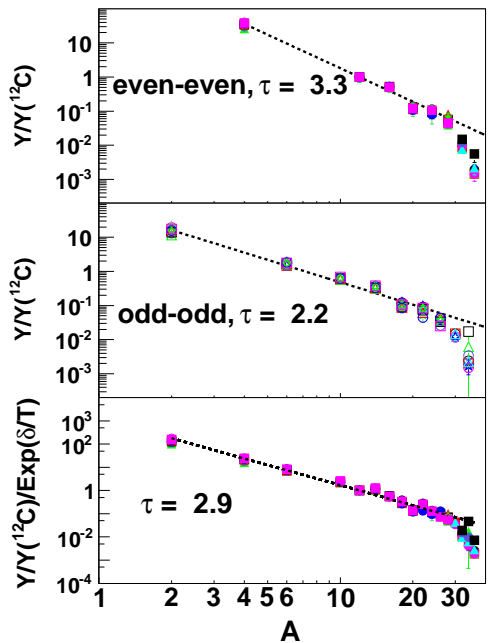}
\caption{\footnotesize (Color online) (Upper) Y/Y($^{12}C$) as a function of A for even-even isotopes with I = 0. (Middle) Same as the upper, but for odd-odd isotopes with I=0. (Bottom) The pairing energy term corrected yields for isotopes with I=0. The lines are the results fitted by $A^{-\tau}$. The extracted $\tau$ values are shown in each figure.
}			
\label{fig:fig2_Exp_RVsA_I0}
\end{figure}
Since the isotope yields of IMFs have been evaluated for the nucleon-nucleon (NN) source component, the same component in the light charged particle emission is also used. This is evaluated in Ref.~\cite{Huang10_1,Chen10}. In order to compare the yields for different reaction systems, all yields are normalized to that of the $^{12}C$ in a given system. 

We study separately the isotope yields for I = 0 and I $\ne$ 0. For the isotopes with I = 0 the symmetry energy contribution in Eq.(\ref{eq:eq_yield}) becomes zero. Since these isotopes can be even-even or odd-odd nucleus, the yields of odd-odd and even-even I = 0 isotopes are plotted separately as a function of A in the top and middle of Fig.\ref{fig:fig2_Exp_RVsA_I0} for the 13 different reactions studied. In each case the distributions from the different reactions are almost identical. 
They show a power law behavior up to A $\sim$ 30. The extracted values of $\tau$ are $\tau$ = 3.3 for even-even and  $\tau$ = 2.2 for odd-odd. The difference in slopes might naturally be attributed to pairing effects. While large pairing effects are expected at low temperatures, because they are related to shell effects~\cite{Preston62}, the disassembling system is initially at a high temperature. Ricciardi et al. have suggested an explanation for the apparent strong effect of pairing in such systems~\cite{Ricciardi04,Ricciardi05}. 
According to their model simulations, experimentally observed pairing effects may be attributed to the last chance particle decay of the excited fragments during cooling. This hypothesis is also supported by our model simulations presented in Ref.~\cite{Huang10_1}. We therefore treat the observed pairing effect as one of the secondary decay effects.

By fitting the yields of even-even and odd-odd isotopes simultaneously and including the pairing coefficient $a_{p}$ in the fitting process we obtain  $\tau = 2.9$ (bottom panel of Fig.\ref{fig:fig2_Exp_RVsA_I0}) and $a_{p}/T = 2.2$.  
Using these parameters, we have divided the normalized yields by the pairing energy contribution, $exp(\delta/T)$, in which $\delta = a_{p}/A^{\frac{1}{2}}$ for even-even, $\delta = 0$ for even-odd and $\delta = - a_{p}/A^{\frac{1}{2}}$ for odd-odd isotopes. The resultant corrected isotope distribution is shown with a fitted line in the bottom figure for all isotopes with I=0.   

\begin{figure}
\includegraphics[scale=0.4]{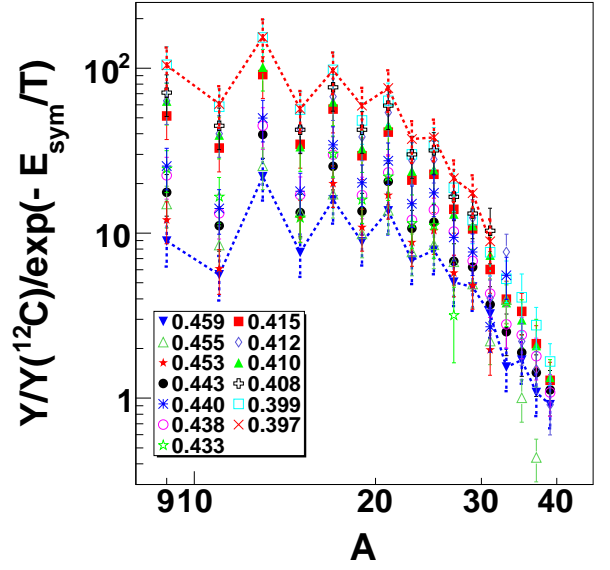}
\caption{\footnotesize 
(Color online) Symmetry term corrected yields of I=3 for different reaction systems. Different symbols present different reactions. Dotted lines are connected between data points for the smallest and largest Z/A values, respectively.  
}			
\label{fig:fig3_RvsA_I3}
\end{figure}

For the isotopes with I $\ne$ 0, one can write the free energy as 
\begin{eqnarray}
F(A,I,T,\rho) = F'(A,I=0,T,\rho) \nonumber\\
- a_{sym}I^2/A + \delta(N,Z).
\label{eq:eq_FEI}
\end{eqnarray}
The formulation indicates that the pairing term for I=0 is excluded and added explicitly into Eq.(\ref{eq:eq_FEI}). 
Since the symmetry contribution is larger for larger I values, we first examine the isotopes with I=3, which is the largest I value for which the yields of a reasonable number of isotope species have been determined. In this case all isotopes are even-odd and therefore the pairing term drops out of Eq.(\ref{eq:eq_FEI}).  
The corrected isotope distributions obtained from the normalized yields divided by $exp(-E_{sym}^{emp}/T)$ are plotted as a function of A in Fig.\ref{fig:fig3_RvsA_I3} for all reactions. Here $E_{sym}^{emp} = a_{sym}^{emp}I^{2}/A$ and $a_{sym}^{emp}$ is given by Eq.(\ref{eq:eq_asymEmp}). One can make a few distinct observations. First, there is a clear even-odd effect. This indicates that the pairing effects can originate, not only from the last chance particle decay, but also from the second-to-last particle decay, the latter in lesser magnitude as discussed in Ref.~\cite{Huang10_1}. In other words, the parents of I=3 isotopes can be I=2 isotopes in the cooling path and the pairing effects are carried on to the I=3 isotopes.

Another observation is a poor scaling between different reactions. Though the distribution does not scale well in magnitude, it is noted that the shapes of the distributions are very similar to each other, especially in the mass range up to $A = 25$. This suggests slight differences of the emitting sources in the different reaction systems. In Eq.(\ref{eq:eq_yield}) for a given isotope, the difference between different reactions comes through the chemical potential terms, $(\mu_{n}N+ \mu_{p}Z)/T$. As pointed out in Ref.~\cite{Huang10_2} the experimental results indicate a relation between isotopic scaling parameters, i.e., $\alpha \sim -\beta $ for these data. This in turn implies the relation, $(\mu_{n} + \mu_{p}) \sim const$. This relation is also suggested using the Quantum Statisitcal Model (QSM) calculation in Ref.~\cite{Botvina02}. Inserting this relation into Eq.(\ref{eq:eq_yield}), one can get 

\begin{eqnarray}
Y(A,I) &\sim& CA^{-\tau}exp\{[(F(A,I,T,\rho)+\mu_{n} I + c Z)/T] \nonumber\\
&&+Nln(N/A)+Zln(Z/A)\},     
\label{eq:eq_yield_I}
\end{eqnarray}
where $\mu_{n} + \mu_{p}=c$. The I dependence of the yield for a given isotope between different reactions, 1 and 2, comes through $\Delta\mu_{n}=\mu_{n}^{1}-\mu_{n}^2$. In the following, we take $\mu_{n}/T = k_1 I ((Z/A)_{sys}-0.5)+\mu_n^0/T$, in which $k_1$ is a parameter determined by minimizing the spread of the data for different I values. $\mu_n^0$ is the chemical potential for symmetric ($N=Z$) systems. By minimizing the spread in Fig.\ref{fig:fig3_RvsA_I3} and those corresponding to the other I values, $ k_1 = -10.3 \pm 0.4$ is obtained. It is worth noting that the k value extracted from the experiment is consistent with the calculated slope of $\mu_{n}$ by the QSM calculation for different N/Z systems, given in Fig.4 of Ref.~\cite{Botvina02}. The QSM calculations also show roughly a linear dependence of $\mu_n$ or $\mu_p$ on N/Z of the system. From that figure, one can get $k_{cal}=slope/T \sim ( \mu_{n}^1(Z/A=0.5) - \mu_{n}^2(Z/A=0.4)/T/((Z/A)^1-(Z/A)^2) \sim -10$ f
 or $T=5$ and $\rho=0.3\rho_0$. (In the figure, the values are given as a function of N/Z.) The calculated slope depends slightly on the temperature and density of the emitting source, i.e., at T=5, $k_{cal} \sim -8$ for $\rho=0.1\rho_0$ and $k_{cal} \sim -12$ for $\rho=0.5\rho_0$.   
   
\begin{figure}
\includegraphics[scale=0.4]{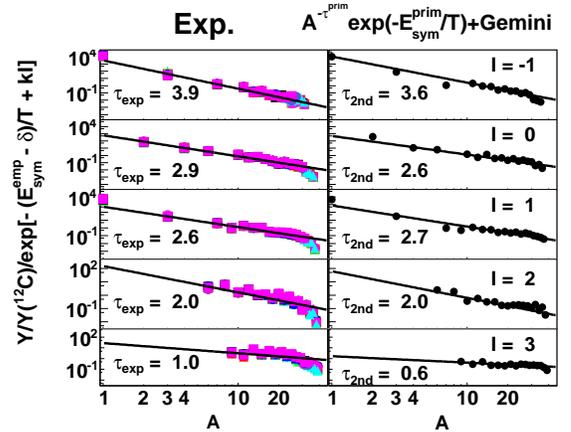}
\caption{\footnotesize (Color online) 
(Left) Corrected experimental isotope distributions for I = -1 to 3 from top to bottom for all 13 reactions. The experimental yields are corrected by $exp(-(E_{sym}-\delta)/T + kI)$, where $k = k_1((Z/A)_{sys}-0.5)$. The pairing term correction, $\delta$, is made only for I=0 and 2. Solid lines are the results by the $A^{-\tau}$ fit for $1 \le A \le 30$. The extracted $\tau$ values are given in each figure. 
(right) Corrected calculated isotope distributions for $\tau^{prim}=2.3$. The same corrections as on the left are made. The extracted $\tau$ values are also given in each figure. 
}	
\label{fig:fig4_RvsA_Im13}
\end{figure} 
 
\begin{figure}
\includegraphics[scale=0.4]{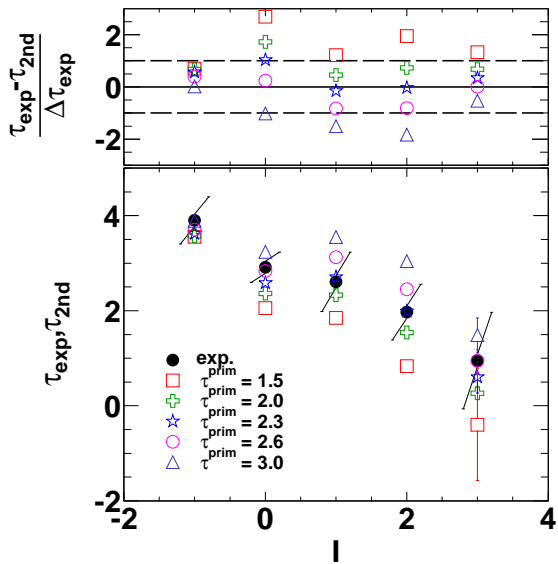}
\caption{\footnotesize
(Color online) Extracted $\tau$ values from the experiment and the final products of the simulations. The experimental results are shown by dots and, for simulations, different symbols represent for those from the simulations with the different $\tau^{prim}$ values, which are indicated in the figure.   
}
\label{fig:fig5_tau_summary}
\end{figure}

The corrected isotope yields are shown in the left column of Fig.\ref{fig:fig4_RvsA_Im13} for I = -1 to 3 from the top to the bottom, including I=0. For all cases the isotope distributions are characterized by a power law distribution, though the spread for I=3 is slightly larger and the quality of the fit decreases. The extracted experimental $\tau$ values, $\tau_{exp}$, decrease systematically from 3.9 to 1.0 as the I value increases from -1 to 3 for the corrected isotope yields. In order to elucidate this observation, a simple de-excitation model simulation was made. The simulation is based on the observation of the power law distribution for I=0 isotopes in Fig.\ref{fig:fig2_Exp_RVsA_I0}. This suggests that the distribution is dominated by the A$^{-\tau}$ term and the A dependence of $F(A,I=0,T,\rho) + \mu_{n}N + \mu_{p}Z$ term is small in Eq.(~\ref{eq:eq_yield}). We extend this assumption to I $\ne$ 0 isotopes, i.e., the primary isotope yields are generated by
\begin{eqnarray}
Y(A,I) &\sim& A^{-\tau}exp\{-a_{sym}^{prim}I^{2}/A/T]\},     
\label{eq:eq_yield_sim}
\end{eqnarray}
where $a_{sym}^{prim}$ is the symmetry energy coefficient of the primary fragments. In the equation the symmetry energy term dependence is kept, although in Fig.\ref{fig:fig4_RvsA_Im13} the symmetry energy term has been corrected using Eq.(\ref{eq:eq_asymEmp}). This is because the value I is not conserved during the de-excitation process and therefore the symmetry energy term corrections, $exp(-E_{sym}^{emp}/T)$ and $exp(-E_{sym}^{prim}/T)$, are made independently. For comparison to these results we have carried out a simple model simulation. In the simulation, we assigned $a_{sym}^{prim}/T = 5$ from Refs.~\cite{Huang10_1,Chen10} and the pairing term is neglected in the primary distribution. We then assume $\tau^{prim}=2.3$ and generate primary isotope yields for $1 \le A \le 50$ and $-2 \le I \le 5$, according to Eq.(\ref{eq:eq_yield_sim}). For each fragment, an excitation energy, Ex, of 3 A MeV is taken~\cite{Marie98,Hudan03}. ( Very similar results are obtained for $ 2.5 \
 \le Ex \le 5.0$ A MeV. For Ex $\le 2.0 $ the extracted $\tau$ value starts to decrease notably.) The statistical de-excitation of these fragments is then followed with the GEMINI code. The intrinsic angular momenta are set to 0 for all IMFs. The same treatment that was applied to the experimental yields has been made for the final product yields obtained with this model. The resultant distributions are plotted in the right side column of Fig.~\ref{fig:fig4_RvsA_Im13}. The extracted $\tau$ values, $\tau_{2nd}$, are given in the figure. The experimental variation of tau values with I are well reproduced by the assumption that $\tau^{prim}=2.3$ for the primary fragments. Different $\tau^{prim}$ values ranging from 1.5 to 3.0
for the primary isotope distribution have also been examined and the results are summarized in the bottom of Fig.~\ref{fig:fig5_tau_summary}. The spread of the experimental values represent the fact that each point for a given A consists of 13 data points. The experimental $\tau$ values are in agreement with those from the simulated events with the primary $\tau$ values in $ 2.0 \le \tau^{prim} \le 2.6$. To determine the best values, the ratio of the defference between $\tau_{exp}$ and $\tau_{2nd}$ to the experimental error is plotted in the top of the figure. From this figure the best fit value for the model is obtained with $\tau^{prim} = 2.4 \pm 0.2$ for the primary fragment distribution.    

In summary, after correction for secondary decay effects the yield distributions for isotopes of different I = N-Z from 13 different reactions in the Fermi energy domain exhibit power law distributions as a function of mass number. The extracted $\tau$ values show a systematic change of the $\tau$ value from 3.9 to 1.0 when the I value of the isotope changes from -1 to 3 and these values are well reproduced by a simple de-excitation model, assuming 
that the isotopic yields of the primary distribution obey a power law dependence with a symmetry term contribution. The experimentally extracted $\tau$ values for each I value are in good agreement with those evaluated from simulations with $\tau_{prim} \sim 2.3$, suggesting that the emitting source of the primary isotopes produced in these reactions is at near the critical point.  

We thank the staff of the Texas A$\&$M Cyclotron facility for their 
support during the experiment. We thank L. Sobotka for letting us to 
use their spherical scattering chamber. We also thank A. Ono and R. 
Charity for letting us to use their calculation 
codes. This work is supported by the U.S. Department of Energy under 
Grant No. DE-FG03-93ER40773 and the Robert A. Welch Foundation under 
Grant A0330. One of us(Z. Chen) also thanks the \textquotedblleft100 
Persons Project" of the Chinese Academy of Sciences for the support.


\begin{thebibliography}{}

\bibitem{Finn82}J. E. Finn {\it et al.}, Phys. Rev. Lett. {\bf 49}, 1321 (1982).
\bibitem{Minich82}R. W. Minich {\it et al.}, Phys. Lett. {\bf B118}, 458 (1982).
\bibitem{Hirsch84}A. S. Hirsch {\it et al.}, Phys. Rev. {\bf C29}, 508 (1984).
\bibitem{Bonasera00}A. Bonasera {\it et al.}, Rivista Nuovo Cimento {\bf 23}, N2 (2000).
\bibitem{Lopez06}O. Lopez {\it et al.}, Eur. Phys. J. {\bf A30}, 263 (2006).
\bibitem{Gilkes94}M. L. Gilkes {\it et al.}, Phys. Rev. Lett. {\bf 73}, 1590 (1994).
\bibitem{Elliott00}J. B. Elliott {\it et al.}, Phys. Rev. {\bf C62}, 064603 (2000).
\bibitem{Elliott02}J. B. Elliott {\it et al.}, Phys. Rev. Lett. {\bf 88}, 042701 (2002).
\bibitem{Elliott03}J. B. Elliott {\it et al.}, Phys. Rev. {\bf C67}, 024609 (2003).
\bibitem{Bonasera08}A. Bonasera {\it et al.}, Phys. Rev. Lett. {\bf 101}, 122702 (2008).
\bibitem{Huang10_3}M. Huang {\it et al.}, Phys. Rev. {\bf C81}, 044618 (2010). 
\bibitem{Marie98}N. Marie {\it et al.}, Phys. Rev. {\bf C58}, 256 (1998).
\bibitem{Staszel01}P. Staszel {\it et al.}, Phys. Rev. {\bf C63}, 064610 (2001).
\bibitem{Hudan03}S. Hudan {\it et al.}, Phys. Rev. {\bf C67}, 064613 (2003)
\bibitem{Tsang01}M. B. Tsang {\it et al.}, Phys. Rev. {\bf C64}, 054615 (2001).
\bibitem{Botvina02}A. S. Botvina, O. V. Lozhkin, and W. Trautmann, Phys. Rev. {\bf C65}, 044610 (2002).
\bibitem{Tsang06}M. B. Tsang, Eur.Phys. J. {\bf A30}, 129 (2006)
\bibitem{Huang10_1}M. Huang {\it et al.}, Phys. Rev. {\bf C81}, 044620 (2010).
\bibitem{Chen10}Z. Chen {\it et al.}, Phys. Rev. {\bf C81}, 064613, (2010)
\bibitem{Ono96}A. Ono and H. Horiuchi, Phys. Rev. {\bf C53}, 2958 (1996).
\bibitem{Ono99}A. Ono, Phys. Rev. {\bf C59}, 853 (1999).
\bibitem{Charity88}R. J. Charity {\it et al.}, Nucl. Phys. {\bf A483}, 371, 1988.
\bibitem{Fisher67}M. E. Fisher, Rep. Prog. Phys. {\bf 30}, 615 (1967).
\bibitem{Preston62}M. A. Preston, Physics of the nucleus, Addison-Wesley Pub. Co. 1962, chapetr 7.
\bibitem{Ricciardi04}M. V. Ricciardi {\it et al.}, Nucl. Phys. {\bf A733} (2004) 299.
\bibitem{Ricciardi05}M. V. Ricciardi {\it et al.}, Nucl. Phys. {\bf A749} (2005) 122c.
\bibitem{Huang10_2}M. Huang {\it et al.}, in press in Nucl. Phys. A, 2010. (arXiv[nucl-ex]:1002.0311, 2010.)

\end{thebibliography}
\end{document}